\documentclass[twocolumn,aps,showpacs,amsmath,amssymb,floatfix]{revtex4}

\usepackage{graphics,color}

\begin{document}

\title{Correlated hybridization in transition metal complexes}

\author{
  A. H\"{u}bsch$^{1,2}$, J. C. Lin$^{1}$, J. Pan$^{1}$, and 
  D. L. Cox$^{1,2,3}$
} 

  \affiliation{
    $^{1}$Department of Physics, University of California, Davis, CA 95616 \\
    $^{2}$Center for Theoretical Biological Physics, University of California,
    San Diego, CA 92119 \\
    $^{3}$Center for Biophotonics Science and Technology, University of
    California, Davis, CA 95616
}

\date{May 15, 2006}

\begin{abstract}
We apply local orbital basis density functional
theory (using SIESTA) coupled with a mapping to the Anderson impurity model to
estimate the Coulomb assisted or correlated hybridization between
transition metal d-orbitals and ligand sp-orbitals for a number of molecular
complexes. We find remarkably high values which can have several physical
implications including: (i) renormalization of effective single band or
multiband Hubbard model parameters for the cuprates and, potentially,
elemental iron, and (ii) spin polarizing molecular transistors.
\end{abstract}


\maketitle

The properties of transition metal compounds are often dominated by the
3\textit{d} orbitals because their localized character causes a strong Coulomb
repulsion between the electrons. The theoretical progress in the field has
been impeded by the extreme difficulties of dealing with even the simplest
model Hamiltonian appropriate for these materials, the Hubbard model
\cite{Hubbard}, consisting of a direct electron hopping between orbitals
\textit{i} and \textit{j} with amplitude $t_{ij}$ and of the Coulomb repulsion
$U$ of electrons in the same orbital. $U$ is a matrix element of the Coulomb
potential, 
$
  U = \langle ii | V(\mathbf{r} - \mathbf{r}')| ii \rangle
$.
Matrix elements involving different lattice sites \textit{i} and \textit{j}
are generally smaller than \textit{U}. One of them is the so-called correlated
hybridization 
$
  X_{ij} = \langle ii | V(\mathbf{r} - \mathbf{r}')| ij \rangle
$
that describes a density dependent hopping. Estimates for $X_{ij}$ ($0.5$~eV
in transition metals \cite{Hubbard}, $0.8$~eV in cuprates \cite{Appel}) show
that the matrix elements of the correlated hybridization are comparable to or
even larger than the corresponding amplitudes of direct hopping in real
systems. Up to now, effects caused by correlated hybridization have not yet
attracted much attention: Hirsch suggested a new mechanism for 
superconductivity \cite{Hirsch}, some exactly solvable cases were discussed
\cite{exact}, finite clusters were studied using exact diagonalization
\cite{Amadon}, metallic ferromagnetism was investigated \cite{Kollar}, and
correlated hybridization was studied via dynamical mean field theory in
the Falikov-Kimball model \cite{Schiller}.

In many compounds, cuprates and manganites are well-known examples, the
transition metal atoms are surrounded by oxygens or other elements with
$p$-orbitals so that the 3\textit{d} orbitals are only effectively
coupled with each other due to oxygen orbitals. Realistic multi-band models
including oxygen degrees of freedom, as proposed 
for the cuprates \cite{Emery}, can not easily be mapped onto
effective single-band models (see \cite{Macridin} and references therein). The
effective hopping between transition metal sites must then be mediated by the
hybridization with $p$-orbitals of the oxygens. Despite the somewhat
robust interest in correlated hybridization between like orbitals, there
have been no {\it ab initio} studies of the corresponding 
{\it correlated hybridization} \cite{karnaukhov} between the $d$ orbitals and 
surrounding $p$ orbitals.  This may be of considerable interest in molecular
transistors based upon transition metal complexes for which the surrounding
ligand atoms provide the linkage to the leads~\cite{Park}.

In this paper, we estimate matrix elements of the correlated hybridization for
several transition metals in different chemical environments from 
density functional theory (DFT) calculations. To our knowledge, this is the
first systematic \textit{ab initio} study of correlated hybridization matrix
elements, and we find remarkably high values, although similar spin dependent
hybridization phenomena have been noted in the context of molecular
transistors \cite{Yu}. We demonstrate
that these additional hybridization matrix elements could 
significantly change the parameters in effective single band models for
transition metal oxides. Furthermore, we demonstrate that the correlated
hybridization may possibly provide a means to significant spin polarization of
currents through  transition metal based molecular transistors in modest
magnetic fields. 

We have carried out spin-polarized electronic structure calculations using 
the fully \textit{ab initio} DFT code SIESTA \cite{SIESTA}. It uses
Troullier-Martins norm-conserving pseudo potentials \cite{Troullier} in the
Kleinman-Bylander form \cite{Kleinman} where we included nonlinear
partial-core corrections for the transition metal atoms to take into account
exchange and corelation effects in the core region \cite{Louie}. We used
the generalized gradient approximation (GGA) for the exchange-correlation
energy functional in the version of Ref.~\cite{Perdew}. SIESTA uses a basis
set of atomic orbitals where the method by Sankey and Niklewski \cite{Sankey}
is employed. We used a double-$\zeta$ basis set, and included
polarization orbitals for the transition metal atoms. To determine the minimum
energy configuration within DFT all complexes were allowed to relay till the
force on each atom was less than 0.03~eV/{\AA}.

The results of the SIESTA calculations are used to determine effective
hybridization matrix elements. For that purpose we map the final DFT 
Hamiltonian $\mathcal{H}_{\mathrm{DFT}}$ obtained from SIESTA onto an
effective two-band model, 
\begin{eqnarray}
  \label{G1}
  \tilde{\mathcal{H}} &=& 
  \sum_{i,\sigma} 
  \tilde{\varepsilon}_{i\sigma}^{d} \, 
  d_{i\sigma}^{\dagger}d_{i\sigma} + 
  \sum_{\alpha,\sigma}
  \tilde{\varepsilon}_{\alpha\sigma}^{p} \,
  p_{\alpha\sigma}^{\dagger}p_{\alpha\sigma} \\
  && +
  \sum_{i,\alpha,\sigma}
  \tilde{V}_{i\alpha\sigma} \,
  \left(
    d_{i\sigma}^{\dagger} p_{\alpha\sigma} + \mathrm{H.c.}
  \right),
  \nonumber
\end{eqnarray}
where the 3\textit{d} orbitals of the transition metal atoms are separated from
the rest. The $d_{i\sigma}^{\dagger}$ ($p_{\alpha\sigma}^{\dagger}$) are
fermionic creation operators of 3\textit{d} (ligand) electrons with spin
$\sigma$ at orbital $i$ ($\alpha$). To determine the one-particle energies, 
$\tilde{\varepsilon}_{i\sigma}^{d}$ and
$\tilde{\varepsilon}_{\alpha\sigma}^{p}$, and the hybridization matrix
elements, $\tilde{V}_{i\alpha\sigma}$, we apply unitary
transformations to $\mathcal{H}_{\mathrm{DFT}}$. Because the atomic basis
states used by SIESTA are 
not orthogonal the transformation of $\mathcal{H}_{\mathrm{DFT}}$ is
performed in three steps: (i) The block of the 3\textit{d} orbitals
(single $\zeta$ only) is diagonalized. (ii) The 3\textit{d} contributions
to the ligand basis states are removed. (iii) The ligand block is
diagonalized. After theses steps we obtain an effective Hamiltonian
$\tilde{\mathcal{H}}$ of the desired form \eqref{G1} where
$\tilde{\mathcal{H}}$ has the same eigenvalues as $\mathcal{H}_{\mathrm{DFT}}$
because $\tilde{\mathcal{H}}$ is derived by unitary transformation from
$\mathcal{H}_{\mathrm{DFT}}$. 

\begin{table}
\caption{
  Maximal difference between the spin directions of the effective hybridization
  matrix elements $\tilde{V}_{i\alpha\sigma}$ for several transition metal
  complexes in the range $\pm 10$ eV about the chemical potential
  where the overlapp between the ligand states of the two spin directions 
  is at least 0.95.
  $A_{\mathrm{max}}$ is defined in Eq.~\eqref{G_Amax}
  and the last column
  gives the orbital symmetry of the maximally different effective
  hybridization matrix elements referenced to tetrahedral or octahedral
  symmetry.
}
\begin{center}
  \scalebox{0.64}{\includegraphics*{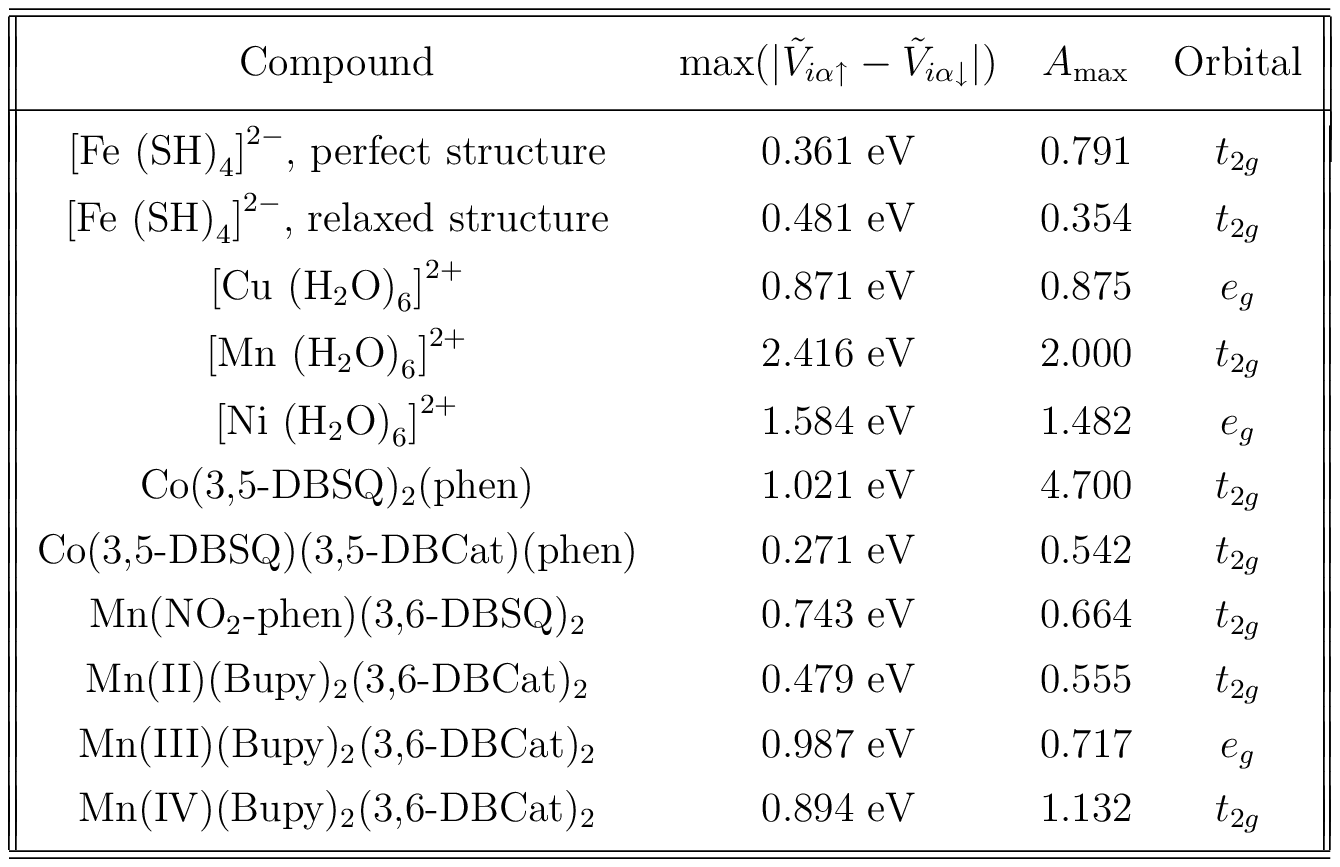}}
\end{center}
\label{table}
\end{table}

Because we performed spin-dependent DFT calculations
$\mathcal{H}_{\mathrm{DFT}}$ consists of two completely separated 
spin sectors. Therefore, the parameters of $\tilde{\mathcal{H}}$ depend on the
spin direction, and we find a remarkable spin dependence of the hybridization
matrix elements if the transition metal atoms are in high-spin states (see
Tab.~\ref{table}). 

\begin{figure}
\begin{center}
  \scalebox{0.81}{\includegraphics*{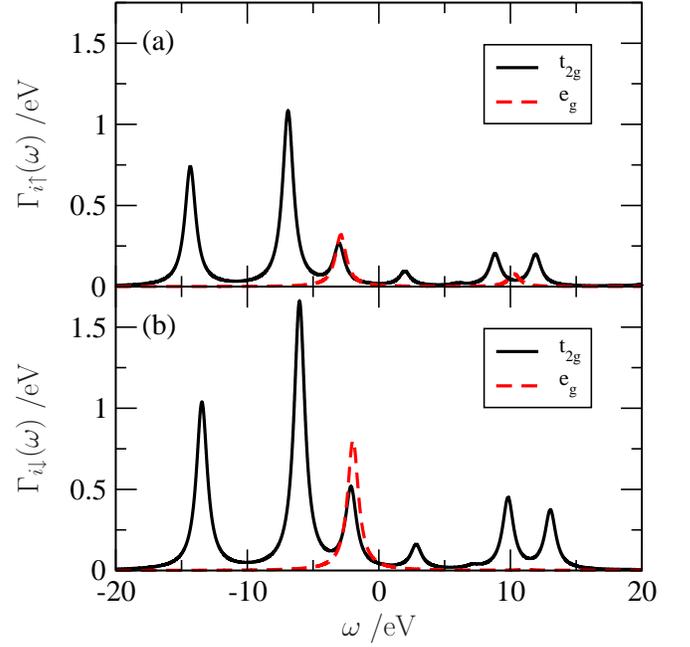}} \\[-5ex]
  $\phantom{a}$
\end{center}
\caption{
  (Color online) Spin dependent hybridization functions for 
  $\left[ \mathrm{Fe} \, \left( \mathrm{SH} \right)_{4} \right]^{2-}$, 
  perfect structure, where the line spectra have been broadened with Gaussian
  functions of width 0.5~eV . The results for the majority [minority] spin
  direction is shown in panel (a) [panel (b)]. 
}
\label{Fig1}
\end{figure}

To discuss the spin dependence of the hybridization matrix elements in 
detail we calculate the hybridization broadening function,
$
  \Gamma_{i\sigma}(\omega) =
  \sum_{\alpha}
  | \tilde{V}_{i\alpha\sigma} |^{2}
  \delta( \omega - \tilde{\varepsilon}_{\alpha\sigma}^{p} )
$,
for all impurity states $i$. As one can see from Fig.~\ref{Fig1}, the
hybridization matrix elements of the minority spin direction are always larger
than the values of the majority spin direction.

A detailed analysis shows that this spin dependence can only derive from
correlated hybridization. To illustrate the point, let us now consider a
Hamiltonian, 
\begin{eqnarray}
  \label{G2}
  \mathcal{H} &=& 
  \sum_{i,\sigma} \varepsilon_{i}^{d} \, 
  d_{i\sigma}^{\dagger}d_{i\sigma} + 
  \sum_{\alpha,\sigma}
  \varepsilon_{\alpha}^{p} \,
  p_{\alpha\sigma}^{\dagger}p_{\alpha\sigma} \\
  && +
  \sum_{i,\alpha,\sigma}
  \left[
    V_{i\alpha} + A_{i\alpha}d_{i-\sigma}^{\dagger}d_{i-\sigma}
  \right]\,
  \left(
    d_{i\sigma}^{\dagger} p_{\alpha\sigma} + \mathrm{H.c.}
  \right),
  \nonumber
\end{eqnarray}
with correlated hybridization matrix elements $A_{i\alpha}$. 
$\mathcal{H}$ can be easily transformed to the form of Eq.~\eqref{G1} if a
factorization approximation is used where the correlated hybridization leads
to spin dependent one-particle energies $\tilde{\varepsilon}_{i\sigma}^{d}$ and
hybridization matrix elements $\tilde{V}_{i\alpha\sigma}$. This
consideration can also be used to determine the matrix elements $A_{i\alpha}$
of the correlated hybridization, one obtains
\begin{eqnarray}
  \label{G3}
  A_{i\alpha} &=& 
  \frac{
    \tilde{V}_{i\alpha\downarrow} - \tilde{V}_{i\alpha\uparrow}
  }{
    \langle d_{i\uparrow}^{\dagger}d_{i\uparrow} \rangle - 
    \langle d_{i\downarrow}^{\dagger}d_{i\downarrow} \rangle
  }.
\end{eqnarray}
Because of the expectation values of the occupation number operators are
restricted, $0 \leq \langle d_{i\sigma}^{\dagger}d_{i\sigma} \rangle \leq 1$,
the difference between the both spin directions of the effective hybridization
matrix elements (as listed in Tab.~\ref{table}) has to be interpreted as a
lower bound for the full matrix element of the correlated 
hybridization. Thus, we conclude from Tab.~\ref{table} that the correlated
hybridization matrix elements $A_{i\alpha}$ are comparable in magnitude to the
regular tight-binding hopping amplitudes $V_{i\alpha}$.
Although the commonly used approximations of DFT such as local density
approximation or CGA are known to fail in accurately describing strongly
correlated transition metal systems (notably the cuprates and metallic
plutonium), there is a notable tradition of successfully using DFT to provide
estimates of parameters for many-body Hamiltonians, as in calculations of the
hybridization for successfully estimating Kondo scales in a series of cerium
heavy fermion systems \cite{Cox}. This is the spirit of our approach; while we
do not include the feedback modification of the density to the DFT, that has
successfully been done only in one instance \cite{Savrasov}.

In the following we want to show that the large values obtained for the
correlated hybridization matrix elements could indeed lead to intriguing
physical effects. For this purpose we consider in the following with the
cuprates and a molecular device two physical systems where the local
coordination environment of the transition metal atoms is precisely the same
as in the clusters studied above. At first, 
we consider a three-band model as proposed for the cuprates \cite{Emery} where
we use standard parameters \cite{param}, 
Cu $3d$ O $2p$ hybridization $t_{pd}=1.3$~eV, 
O $2p$  O $2p$ hybridization $t_{pp}=0.65$~eV, 
Cu on-site repulsion $U_d=8.8$~eV, and
charge transfer energy $\Delta=3.5$~eV. 
Furthermore, we add a correlated hybridization between Cu $3d$ and O $2p$
orbitals to the Hamiltonian with matrix elements $A_{pd}=0.5$~eV. In
accordance to our DFT estimates, $A_{pd}$ has the same sign and phase
factors as $t_{pd}$.  The three-band model can be reduced \cite{eff_J} to a
$S=\frac{1}{2}$ Heisenberg model on the square lattice of Cu sites where the
coupling strength in fourth order perturbation theory is given by 
$
  J =
  \frac{4 t_{pd}^{4}}{2\Delta^{3}} + 
  \frac{4 t_{pd}^{2} (t_{pd} + A_{pd})^{2}}{\Delta^{2}U_d}
$
Thus, the matrix elements $A_{pd} = 0.5$~eV of the correlated hybridization
lead to an increase of the Heisenberg exchange from 0.24~eV to 0.34~eV.

The correlated hybridization also affects the mapping of the three-band
model onto an effective single-band model by means of a cell perturbation 
theory \cite{Macridin, cell_PT}. Despite an increase of the magnitudes of the
effective hopping matrix elements by about 20\%, the on-site Hubbard $U$ is
significantly lowered from $2.48$~eV to $1.94$~eV. This reduction
in $U$ may have some relevance in understanding the need to use smaller
Hubbard interactions in models of, e.g., metallic Fe than one obtains from
estimates using constrained occupancy DFT \cite{Anisimov}:  
If there is significant correlated hybridization of $d$ levels with $s,p$
levels on a neighboring site then the same mechanism of $U$ reduction upon
folding down to the $d$-band only model from the multi-band model will be
operable. 

\begin{figure}
\begin{center}
  \scalebox{0.73}{\includegraphics*{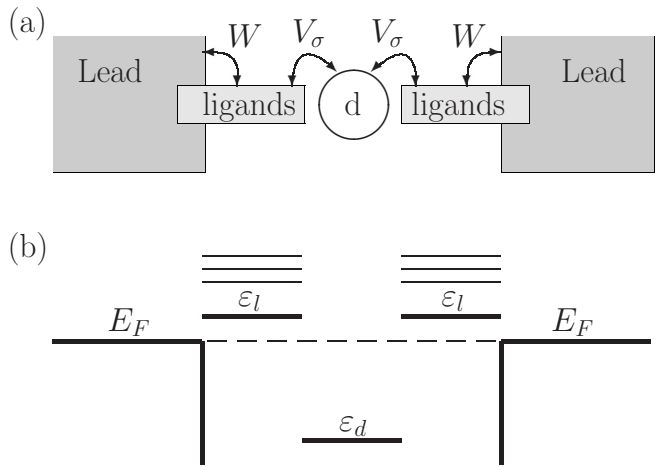}}
\end{center}
\caption{
  Panel (a) shows the schematic design of the proposed spin filter device with
  the definitions of the lead-to-ligand and ligand-to-transition metal site
  hybridization matrix elements $W$ and $V_{\sigma}$. The energies of the
  involved electronic states are sketched in panel (b).
}
\label{Fig2}
\end{figure}

Perhaps the most interesting application of our results is to tunable molecular
transistors \cite{Park,transport,transistor}, with the strongest response
possible not in the Kondo limit but rather the mixed valent limit.  In the
presence of a relatively modest field, the correlated hybridization can impart
a potentially significant spin polarization of the current through the device
without requiring spin polarization of the leads as has been considered both
experimentally \cite{Ralph} and theoretically \cite{theo_fmlead}. To
illustrate the idea, consider the schematic device of Fig.~\ref{Fig2}, in
which a transition metal based molecule is attached to ordinary metallic
leads, and assume that a spin $1/2$ large $U$ model is applicable, as relevant
for the low spin Co complexes considered in Ref.~\cite{Park} (electron
hopping) or a Cu complex (hole hopping).  The bare lead-to-transition metal
site hybridization $V_{Ld}$ goes as 
$ V_{Ld}^{\sigma} \approx \frac{W \, V_{\sigma}}{\varepsilon_{l}-E_{F}} $
where $W$ is the lead-to-ligand hybridization matrix element, $\varepsilon_L$
is the ligand LUMO (HOMO) for electrons (holes) to transition from lead to
ligand, and $V_\sigma=V_0(1-A m \sigma)$ is the spin dependent hybridization
discussed earlier, with $A$ a dimensionless ratio of the assisted hopping
matrix element to the direct one $V_0$, and $m$ the induced polarization
$m=n_{d\uparrow}-n_{d\downarrow}$ of the transition metal ion in an applied
field $H$. Note that estimates,
\begin{eqnarray}
  \label{G_Amax}
  A_{\mathrm{max}} &=&
  2 \,
  \frac{\max|\tilde{V}_{i\alpha\downarrow} - \tilde{V}_{i\alpha\uparrow}|}
  {\tilde{V}_{i\alpha\downarrow} + \tilde{V}_{i\alpha\uparrow}},
\end{eqnarray}
for the ratio $A$ can be found in Tab.~\ref{table}.  Importantly, we note that
$A$ can be as large as 3-5. 

We next assume that the Friedel sum rule \cite{Hewson} can be applied so that
the number of electrons (holes) $n_d$ on the transition metal site is given by 
$
  n_d = n_{d\uparrow}+n_{d\downarrow} = 
  \frac{\delta_{\uparrow}+\delta_{\downarrow}} {\pi} 
$
and the polarization $m$ by 
$
  m = \frac{1}{\pi}(\delta_{\uparrow}-\delta_{\downarrow})
$
where $\delta_{\sigma}$ is the Fermi energy phase shift for an electron of
spin $\sigma$ to scatter off the transition metal site. We further take a
renormalized resonant level model for the phase shift, with resonance position
$\tilde\varepsilon_{\sigma} = \tilde\varepsilon_0-\mu H\sigma$, $\mu$ the
magnetic moment of the d-levels, and resonance width 
$\tilde\Gamma_{\sigma} = \tilde\Gamma_0(1-2 A m\sigma)$ where only
contributions linear in the magnetic field $H$ are taken into account. Note
that in the Kondo regime the zero field resonance parameters satisfy 
$\sqrt{\tilde\varepsilon_0^2+\tilde\Gamma_0^2}\approx k_BT_K$, with $T_K$
the Kondo temperature of the transistor. In this model, the phase shift is 
$
  \delta_{\sigma} = 
  \frac{\pi}{2} - 
  \tan^{-1}\left(
    \frac{\tilde\varepsilon_{\sigma}}{\tilde\Gamma_{\sigma}}
  \right)
$.
Assuming zero bias, identical leads, and low temperature, the
conductance $\cal{G}_{\sigma}$ within our mean-field treatment of the
correlated hybridization in spin channel $\sigma$ is approximately given
by \cite{Wingreen}, 
$\mathcal{G}_{\sigma} \approx \frac{{e}^{2}}{\pi\hbar} \sin^2\delta_{\sigma}$.

Using the definitions and results of the previous paragraph we thus find  
\begin{eqnarray}
  \label{G4}
  \frac{(\mathcal{G}_{\uparrow}-\mathcal{G}_{\downarrow})}
  {\mathcal{G}(H=0)} 
  &\approx&
  \frac{2\sin(2\delta_0)}
 {1+ \frac{2A}{\pi}\sin(2\delta_0)}\frac{\mu H}{\tilde\Gamma_0} 
\end{eqnarray}
with $\delta_0$ the zero field phase shift.  

The $A$ dependent enhancement factor follows from the impurity
polarization $m \propto {\cal G}_{\downarrow}-{\cal G}_{\uparrow}$.  This
effect is overestimated within our mean field approximation, although it has a
clear physical origin:   the up-spin resonance narrows significantly in
applied field as the down spin resonance broadens, allowing a feedback effect
that enhances the tendency towards magnetization saturation.
The effect is most pronounced in the mixed valent regime for $n_d\approx
1.5$ where a critical $A$ value of $\pi/2$ yields divergent $m$
within the approximation. Indeed, a significant enhancement of the zero field
local susceptibility was found in a nonperturbative treatment of this model
for $U_d=0$ in the mixed valent regime \cite{karnaukhov}. It might be possible
to tune the resonance into the highly polarizing regime through a combination
of gate control (to tune $\epsilon_d$) and contact chemistry (to tune the
lead-transition metal ion hybridization). The enhancement effect can also be
significant in the Kondo regime, for which $|1-n_d|\le 0.3$ \cite{bcw} and
$|\sin(2\delta_0)|<0.81$.   
Examining the $A$ values in our table, we obtain critical $|1-n_d|$ values of
0.11 and 0.29 for the Co(II)(3,5-DBSQ)$_2$(phen) and [Mn(H$_2$O)$_6$]$^{2+}$
complexes, safely within the Kondo regime. 

To conclude, we have presented here the first systematic \textit{ab initio}
study of correlated hybridization matrix elements for several transition metal
complexes. Based on DFT calculations we have found correlated
hybridization matrix elements comparable in magnitude to the regular
tight-binding hybridization amplitudes which can lead to significant changes
in the parameters of effective models for transition metal compounds. Finally,
we have sketched how the correlated hybridization can be employed to design
spin sensitive devices, polarizing the spin current dramatically with
potentially modest applied fields and paramagnetic leads.

We acknowledge helpful discussions with T.M. Bryant, S.Y. Savrasov,
A. Schiller, and R.R.P. Singh. This work was supported by NSF grant PHY
0120999 (the Center for Biophotonics Science and Technology), by the US
Department of Energy, Division of Materials Research, Office of Basic Energy
Science, and by DFG through grant HU 993/1-1. DLC is grateful for the support
of the Center for Theoretical Biological Physics through NSF grants PHY0216576
and PHY0225630, and the J.S. Guggenheim Memorial Foundation.


\end{document}